\documentclass[superscriptaddress,twocolumn,10pt]{revtex4-1}

\usepackage[dvips]{graphicx}
\usepackage{latexsym}
\usepackage{amsmath}
\usepackage{amssymb}
\usepackage{amsfonts}
\usepackage{color}
\usepackage{bm}
\usepackage{verbatim}
\usepackage[english]{babel}
\usepackage[utf8]{inputenc}
\usepackage[T2A]{fontenc}
\usepackage{pdfpages}
\usepackage{subfigure}
\usepackage{wrapfig}

\renewcommand{\Re}{\mathop{\mathrm{Re}}}
\renewcommand{\Im}{\mathop{\mathrm{Im}}}
\newcommand{\sign}{\mathop{\mathrm{sign}}}

\begin{document}

\title{
	Electronic structure of vortices pinned by columnar defects in $p_x \pm
	i p_y$ superconductors. 
}

\author{V. L. Vadimov}
\affiliation{Institute for Physics of Microstructures, Russian
Academy of Sciences, 603950 Nizhny Novgorod, GSP-105, Russia }
\affiliation{Lobachevsky State University of Nizhny Novgorod, 23 Prospekt Gagarina, 603950, Nizhny Novgorod, Russia}
\author{A. S. Mel'nikov}
\affiliation{Institute for Physics of Microstructures, Russian
Academy of Sciences, 603950 Nizhny Novgorod, GSP-105, Russia }
\affiliation{Lobachevsky State University of Nizhny Novgorod, 23 Prospekt Gagarina, 603950, Nizhny Novgorod, Russia}

\begin{abstract}
	The electronic structure of a vortex pinned by an insulating columnar
	inclusion in a type-II chiral $p_x \pm i p_y$ superconductor is studied
	within the  Bogolubov-de Gennes theory. The structure of the anomalous
	spectral branch is shown to be strongly affected by the mutual orientations of
	the angular momenta of the center of mass and the relative motion of the two electrons in the Cooper pair.
	Being only slightly perturbed by the scattering at the defect for the zero sum of these angular momenta
	the anomalous spectral branch appears to change dramatically in the absence of such compensation.
	In the latter case the defect presence
	changes the anomalous branch slope sign at the Fermi level
	resulting in the quasiparticle angular momenta inversion at the positive
	energies and the impact parameters smaller than the
	defect radius.
	The experimentally observable consequences for the scanning tunneling
	microscopy characteristics and high-frequency field response are discussed.
\end{abstract}

\maketitle
\section{Introduction}

The experimental search for superconducting compounds with unconventional order parameters
remains a "hot topic" in the condensed matter community during the recent decades.
Such activity is certainly accompanied by a lot of theoretical works aimed to suggest reliable tests probing
the gap anisotropy at the Fermi surface. These tests are usually based either on the gap nodes presence at the Fermi surface
or on the order parameter peculiar phase structure in the momentum space. The latter group of suggestions are especially
effective for the detection of the so -- called chiral superconductivity which
is proposed to be realized, e.g., in
$\mathrm{Sr}_2\mathrm{Ru}\mathrm{O}_4$ \cite{refMackenzieMaenoRevModPhys, refNelsonMaoScience, refXiaMaenoPhysRevLett} or
heavy-fermions compounds \cite{refFiskHessScience,
refGannonShivaramEurophysLett}.

In particular, the nontrivial phase structure of the gap in the momentum space
can revealed by an external magnetic field
which introduces vortex lines in the superconducting sample. The resulting
inhomogeneous superconducting state can reveal
many unusual magnetic and transport properties originated from the interplay
between the nonzero vorticities in both the momentum and
coordinate spaces. This interplay is known to be responsible for specific
structure of quasiparticle subgap states inside the vortex cores
investigated, e.g., in Ref~\cite{refVolovikMajorana}. Unfortunately, the key difference in the
quasiparticle spectra of the vortex core states from the standard
Caroli--de Gennes--Matricon (CdGM) ones appears only beyond the quasiclassical
consideration. Namely, the anisotropic gap structure can shift the quasiparticle
energy quantization rules causing the changes in the minigap value. Taking
the simplest gap in the form $p_x \pm i p_y$ we can get a complete suppression
of the minigap \cite{refVolovikMajorana}. The resulting zero energy mode appears to be extremely robust
to the perturbations such as the ones caused by scattering centers,
etc. \cite{refVolovikMajorana}. However,
the interlevel energy distances inside the cores are the order of
$\Delta_0^2/E_F$, where $\Delta_0$ is the superconducting gap magnitude and $E_F$
is Fermi energy, which is very small value and the
observation of the levels demands high resolution experimental techniques. The existing
scanning tunneling microscopy (STM)/scanning tunneling spectroscopy (STS) data, for
instance, can not provide the reliable evidence for the minigap existence.

In this paper we suggest an alternative way to detect the chiral superconductivity
in the vortex state studying the distinctive features of the vortex electronic
structure in the presence of rather large columnar defects parallel to the applied magnetic
field so that vortex is pinned over the entire length.
Such defects can be created artificially by proton or heavy
ion irradiation, by normal particles and nanorods inclusion and by introducing
arrays of submicrometer holes. The quasiparticle spectrum in the vortex
trapped by the columnar defect was studied in the
Refs.~\cite{MelnikovSamokhvalovPhysRevB79p134529} and~\cite{refRosensteinShapiroDeutchPhysRevB84134521} for the simplest case of
conventional $s$-wave superconductor. It was shown that the defect leads to
significant minigap increase up to the values of the order $\Delta_0$. One can
expect that for unconventional superconductors the physical picture of the defect influence
on the vortex states
can become much more complicated and spectacular since in this case the subgap states bound
to the defects or the sample surface can appear even
without any vortices \cite{refHuPhysRevLett721526, refTanakaPhysRevLett743451}.
These edge states should interact with the vortex bound states changing, thus, the resulting spectrum significantly.
 The hybridization of these quasiparticle modes has been studied previously
 for a particular case of a mesoscopic disc trapping a
singly-quantized vortex \cite{refSilaevJETP87}. It was shown that the vortex and
antivortex spectra differ qualitatively due to the interplay between the internal
vorticity in the momentum space and the vorticity in the coordinate space
associated with the vortex.

Besides its fundamental interest
the problem of pinned vortex spectrum in type-II superconductors with anisotropic gap
is particularly important for understanding the nature of dissipation
 in such compounds. Indeed, vortex pinning has an
influence on the vortex motion and, thus, strongly affects the superconductor
transport properties in the flux flow regime.
The microscopic consideration of pinned vortex matter should become important, of course,
either at low temperatures well below the critical one or for defect dimensions smaller than the coherence length $\xi$
(see discussion in
Refs.~\cite{refThunebergKurkijarviPRL,refThunebergKurkijarviPRB,
refThunebergLTP,refMelnikovSamokhvalovJETPTopoTransition}).
The opposite limits corresponding to the temperature range close to the critical temperature
or large defects can be perfectly described
 within phenomenological
approaches \cite{refMkrtchyan,refNordborgPhysRevB6212408,refBuzdinFeinberg,
refBuzdinDaumens,refBespalovMelnikov}.


\begin{figure}[h]
	\centering
	\includegraphics[width=0.9\columnwidth]{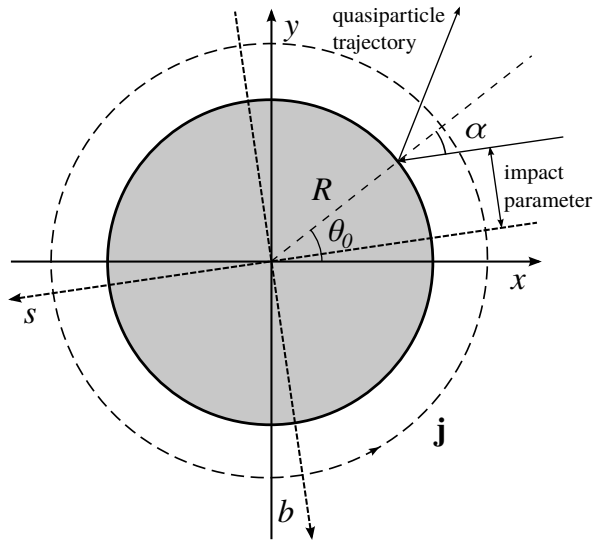}
	\caption{Quasiparticle specular reflection at the defect surface. Here $s$ and $b$ are the coordinates
chosen along and perpendicular to the classical trajectory.}
	\label{figTrajectory}
\end{figure}

In this paper we do not consider, of course, the full problem of collective pinning which can be resolved only
taking account of the vortex -- vortex interaction. Instead we focus on the consideration of the individual
pinned vortex and
analyze the modification of the quasiparticle spectrum  caused by the columnar defect assumed to
have the form of an insulating cylinder of a finite radius $R <\xi$(see Fig.
\ref{figTrajectory}).
We study the transformation of the anomalous energy branches
originated from the normal reflection at the defect boundary.  Let us elucidate the key point of our work and start
from a qualitative discussion of the spectrum transformation.
We analyze the spectrum within the one-dimensional
quasiclassical quantum mechanics of the electrons and holes propagating along
the classical
trajectories. Each trajectory is defined by the impact parameter $b$ and the
trajectory orientation angle $\theta_k$. We assume the reflection to be specular and the trajectory
experiencing the break at the defect consists of two rays with the same
impact parameters and different $\theta_k$ values satisfying
the Snellius law. 
According to the general theory \cite{refBeenakkerVanHoutenPRL} the quasiparticle spectrum at each trajectory is determined by
 the order parameter phase difference $\delta \varphi$ at the ends of the trajectory. This
phase difference has two contributions arising from
 the vorticities in the coordinate and momentum spaces:
$\delta\varphi = \delta\varphi_{r}+\delta\varphi_{k} $.
The first term
$\delta\varphi_{r} = \mp 2 \arcsin (b / R)$ originates from the vortex phase in the coordinate
space and the signes $-$ and $+$ correspond to opposite vorticities.
The second term
 results from the order parameter phase circulation
in the momentum space and for the simplest choice of the chiral $p$-wave gap function
   $\Delta \propto e^{i \theta_k}$ this contribution takes the form:
   $\delta\varphi_{k} = -2\arcsin (b / R) - \pi$.
Using the standard expression $\epsilon_J(b) = -\Delta_0 \cos(\delta \varphi /
2) \sign [\sin (\delta \varphi / 2)]$ for the subgap quasiparticle energy level in a single mode Josephson junction
\cite{refBeenakkerVanHoutenPRL} we find:
$\epsilon_J(b) = - 2 \Delta_0\sign
\left(1-2b^2/R^2\right)b / R
\sqrt{1 - b^2/R^2}$ and
$\epsilon_J(b) = 0$ for the coinciding and opposite vorticities in the
coordinate and the momentum spaces, respectively. Let us denote these vortices
as $N_+$ vortex and $N_-$ vortex, correspondingly.
The above result illustrates the qualitative
difference between the chiral and conventional superconductors. For the latter
case the total phase difference $\delta \varphi$ is completely determined by $\delta
\varphi_r$ and, thus, the change of vorticity sign cannot change the energy levels.

The quasiparticles with large impact parameters $|b| > R$ do not experience the
reflection at the defect surface, so the spectrum in this case should be the
same as the CdGM spectrum. Provided this crossover occurs at small $b\ll\xi$
we get $\epsilon_{CdGM}(b) \approx \Delta_0 b / \xi$ while in the large $b\gtrsim\xi$ limit
the spectrum saturates at the gap value \cite{refCdGM}.
The above reasoning cannot give this crossover
to the CdGM branch because the Doppler shift associated with the superfluid
velocity has been omitted.
One can expect that in the limit $R\ll\xi$ and $b<R$
this Doppler shift contribution can not exceed the value of the CdGM energy $\sim\Delta_0 b/\xi$.
Thus, for $N_-$ vortices the spectrum should only slightly
deviate from the Fermi level being close to the CdGM solution. In the case of $N_+$
vortices the Doppler shift correction is small comparing to the
term $-2\Delta_0 b/R$ introduced above.
The negative sign in the latter expression results in the important
spectrum peculiarity: the inversion of the anomalous branch slope
for the $N_+$ vortex pinned by the defect comparing to the slope for a free vortex.
Note that this inversion effect
has been overlooked in numerical calculations presented in Ref.~\cite{refRosensteinShapiroJOP}.
The change in the slope sign can cause rather drastic changes in the measurable characteristics.
In particular, according to the
spectral flow theory \cite{refKopninBook} it is the behavior of the anomalous
branch which determines the high-frequency conductivity and Hall effect.
In this paper we discuss possible influence
of such spectrum transformation on the transport properties of the
vortex state at finite frequencies.

The rest of the paper is organized as follows.
In Section~\ref{secModel} the basic
equations used for the spectrum calculation are introduced. In Sec.~
\ref{secExcitationSpectrum} we find the quasiparticle spectrum for a
single-quantum vortex trapped by a columnar defect. The Sec.~\ref{secLDOS}
is devoted to the analysis of the local density of states (DOS). In the next
Sec.~\ref{secHallConductivity} we discuss the defect influence on the
high-frequency field response. The results are summarized in the last Sec.~\ref{secSummary}.

\section{Model}
\label{secModel}

We consider a columnar defect as an insulator cylinder of the radius $R$. The magnetic field
$\mathbf B$ is parallel to the cylinder axis $z$, the vortex axis coincides with the cylinder axis.
Thus, the system is invariant with respect to the translations along the $z$-axis and the rotations
around it. For simplicity we restrict ourselves with a two-dimensional case
and consider a motion of quasiparticles only in  the $(x, y)$ plane. The excitation
spectrum can be obtained from the Bogolubov-de-Gennes equations (BdG) written for a two-component
quasiparticle wave function $\psi(\mathbf{r}) = (u,\;v)$:
\begin{equation}
	\label{eqBdGExact}
	-\frac{\hbar^2}{2m}\left(\nabla^2 + k_F^2\right) \tau_3 \psi +
	\left(\widehat \Delta(\mathbf{r}) \tau_+ + h.c.\right)\psi
	= \epsilon \psi \ ,
\end{equation}
where $\tau_\pm = (\tau_1 \pm i \tau_2) / 2$, $\tau_1$, $\tau_2$ and
$\tau_3$ are the Pauli
matrices in the Nambu space, $\hbar k_F$ is the Fermi momentum, $\widehat \Delta$ is
the superconducting gap operator. Considering an extreme type-II
superconductor with a large London penetration depth  $\lambda\gg \xi$
we neglect the vector potential of
the magnetic field $A_\theta \approx B r / 2$ because its contribution to the
superfluid velocity $A/\Phi_0
\propto r/\lambda^2$ is small
compared to the gradient of the order parameter phase $\propto
1/r$\cite{refKettersonSong}. Here $r$
and $\theta$ are the polar coordinates and $\Phi_0$ is the magnetic flux
quantum.
We assume that the quasiparticle wavefunction does not penetrate the defect and
imply the zero boundary conditions at the defect surface:
\begin{equation}
	\label{eqBoundaryExact}
	\psi(R, \theta) = 0 \ .
\end{equation}

\subsection{Quasiclassical approach}

The superconducting gap $\widehat \Delta$ varies at the spatial scale $\xi$
which is much greater than the atomic scale $k_F^{-1}$ in all known
superconductors. Hence one can solve the system~(\ref{eqBdGExact}) within the
quasiclassical approximation. We follow the
approach described in Refs.~\cite{refKopninBook,MelnikovSamokhvalovPhysRevB79p134529} and introduce
the momentum representation:
\begin{equation}
	\label{eqMomentumRepresentation}
	\psi(\mathbf r) = \int \frac{d^2k}{(2\pi)^2} e^{i\mathbf{kr}}\psi(\mathbf k) \ ,
\end{equation}
where $\mathbf k = k (\cos \theta_k, \sin \theta_k) = k \mathbf{k}_0$. The
unit vector $\mathbf k_0$ which depends on the angle $\theta_k$ determines the
trajectory direction in the $(x,y)$ plane. We assume that the solutions of the
equation~(\ref{eqBdGExact}) correspond to the absolute momentum values close
to  $\hbar k_F$:
$k = k_F + q$ ($q \ll k_F$). Using the Fourier transformation
\begin{equation}
	\psi(\mathbf k) = \frac{1}{k_F} \int\limits_{-\infty}^{+\infty}
	ds\;e^{i(k_F - k) s}g(s, \theta_k) \ .
\end{equation}
One can finally express the wavefunction in the coordinate space
$\psi(\mathbf r)$ through the function $g(s, \theta_k)$:
\begin{equation}
	\label{eqQuasiClassicSubstitute}
	\psi(r, \theta) = \int\limits_0^{2\pi} e^{i k_F r \cos(\theta_k - \theta)}
	g \left[ r \cos(\theta_k - \theta), \theta_k\right]\; \frac{d\theta_k}{2\pi} \ .
\end{equation}
To obtain the quasiclassical equation along the quasiparticle
trajectory we inroduce an angular eikonal:
\begin{equation}
	\label{eqAngularEikonal}
	g(s,\theta_k) = e^{i S(\theta_k)} \overline \psi(s, \theta_k) \ ,
\end{equation}
assuming $\overline \psi$ to be a slowly varying function of $\theta_k$.
Quasiparticles propagating along the trajectories are characterized by the
angular momentum $\mu$:
\begin{equation}
	\mu = -k_F b = \frac{\partial S}{\partial \theta_k} \ .
\end{equation}
Here we introduce the impact parameter $b$ which determines the classical
trajectory along with the $\theta_k$ angle. The angular momentum is conserved
due to the axial symmetry of the system.

Substituting~(\ref{eqQuasiClassicSubstitute}) and~(\ref{eqAngularEikonal}) into~(\ref{eqBdGExact}) one can
obtain the equation for $\overline \psi$ at the
classical trajectory~(see Fig.~\ref{figTrajectory}):
\begin{equation}
	\label{eqBdGQuasiClassic}
	-i\hbar v_F \tau_3 \frac{\partial \overline \psi}{\partial s} +
	\left(\tau_+ \overline \Delta(\mathbf r, \theta_k) + h.c. \right)
	\overline \psi = \epsilon \overline \psi \ ,
\end{equation}
where $m v_F = \hbar k_F$, $\overline \Delta$ is the quasiclassical form of
the gap operator, and $s$ is a coordinate along the classical trajectory.
The transition from  the $(s, b)$ coordinates~(see Fig.~\ref{figTrajectory})
to the usual cartesian $(x, y)$
coordinates can be performed as follows:
\begin{equation}
	\label{eqTrajectoryCoords}
	x = s \cos \theta_k - b \sin \theta_k,~ y = s \sin \theta_k + b \cos
	\theta_k \ .
\end{equation}
Further consideration requires an explicit expression of the gap operator $\widehat \Delta$.

In the homogeneous case the gap operator for the superconductors with spin-triplet coupling is
determined by a momentum -- dependent vector $\mathbf d$ \cite{refSigristUedaRevModPhys}:
\begin{equation}
	\widehat\Delta(\mathbf k) = -i\sigma_y (\mathbf d(\mathbf k), \mathbf \sigma)
	= \begin{pmatrix}
		-d_x(\mathbf k) - i d_y(\mathbf k)	&	d_z(\mathbf k)\\
		d_z(\mathbf k)				&	d_x(\mathbf k) - i d_y(\mathbf k)\\
	\end{pmatrix} \ .
\end{equation}
We consider a superconductor with $d_x = d_y = 0$ and $d_z \propto \hat k_x +
i \hat k_y,~ \hat k_x - i \hat k_y$. Such an order parameter possibly
describes a superconducting state in $\mathrm{Sr_2RuO_4}$ \cite{refRiceSigrist} and heavy-fermion
compounds \cite{refSungkitAnupamPRB}. Separating the spin dependence and generalizing the operator for
the inhomogeneous case we get the following expression for $\Delta$:
\begin{equation}
	\widehat \Delta = \frac{\Delta_0}{2 k_F}\left(\left\{\eta_+(\mathbf r),\;-i\partial_+
	\right\} + \left\{\eta_-(\mathbf r),\;-i\partial_-
	\right\}\right) \ ,
\end{equation}
where $\Delta_0$ is a gap magnitude, $\eta_\pm(\mathbf r)$ are the coordinate
depending order parameters, which describe the Cooper pairs with the
opposite angular momenta directions, $\{a, b\} = ab + ba$ is an
anticommutator, $\partial_\pm = \partial/\partial x \pm i \partial / \partial y$.
Two degenerate ground states are described by the
following  order parameters:
$\eta_+ = 1$, $\eta_- = 0$ and $\eta_- = 1$, $\eta_+ = 0$. In a general inhomogeneous case
both order parameters are non zero while usually far from the
topological defects one of them is suppressed. The areas where only one order
parameter is nonzero are called chiral domains.

One can find the quasiclassical form of $\widehat \Delta$, neglecting the terms
of the order of $(k_F \xi)^{-1}$:
\begin{equation}
	\overline \Delta(\mathbf r, \theta_k) = \Delta_0 \left(\eta_+(\mathbf r) e^{i \theta_k}
	+ \eta_-(\mathbf r) e^{-i \theta_k}\right) \ .
\end{equation}
Hence the equation~(\ref{eqBdGQuasiClassic}) takes the following form:
\begin{equation}
	\label{eqBdGPType}
	-i \xi \tau_3 \frac{\partial \overline \psi}{\partial s} + \left(D(\mathbf
	r, \theta_k) \tau_+ + h.c.\right)
	\overline \psi = \varepsilon \overline \psi \ ,
\end{equation}
where $\xi = \hbar v_F / \Delta_0$ is the coherence length, $D(\mathbf r,
\theta_k) = \eta_+(\mathbf r)\exp(i \theta_k) +
\eta_-(\mathbf r) \exp(-i \theta_k)$, $\varepsilon = \epsilon / \Delta_0$.

The axially symmetric vortex solutions are described by the following order parameters
\cite{BarashMelnikovZhETFEn}:
\begin{equation}
	\label{eqPOrderParameterAnsatz}
	\eta_\pm(\mathbf r) = f_\pm^{(m)}(r)e^{i(m \mp 1)\theta} \ ,
\end{equation}
where $m$ is the sum of the winding numbers in the coordinate and the momentum
spaces. One of the functions $f_\pm$ saturates at unity at large $r
\gg \xi$ and another one vanishes far from the core.
The magnetic flux carried by the vortex is determined
by the winding number of the dominating order parameter component,
i.e. it is equal to
$m + 1$ flux quanta for a vortex trapped in the $\eta_-$ domain and $m - 1$ flux
quanta for a vortex in the
$\eta_+$ domain. Using~(\ref{eqTrajectoryCoords}) and (\ref{eqPOrderParameterAnsatz})
we find the order parameter profile at
the classical trajectory:
\begin{equation}
	D_b(s, \theta_k) = e^{i m \theta_k} \sum
	f_\pm^{(m)}\left(\sqrt{s^2 + b^2}\right) \left(\frac{s + ib}{\sqrt{s^2 + b^2}} \right)^{m \mp 1} \ .
\end{equation}
We can separate the $\theta_k$ dependence, introducing a new function:
\begin{equation}
	\label{eqMomentumSubstitute}
	\overline \psi = e^{i (m \tau_3/2)\theta_k} \widetilde \psi
\end{equation}
and the equation~(\ref{eqBdGPType}) becomes:
\begin{equation}
	\label{eqBdGPType1}
	-i \xi \tau_3 \frac{\partial \widetilde \psi}{\partial s} + \left(\widetilde D_b(s) \tau_+ + h.c.\right)
	\widetilde \psi = \varepsilon \widetilde\psi
\end{equation}
where $\widetilde D_b(s) = \exp(-im\theta_k) D_b(s, \theta_k)$. For definiteness
we consider the $\eta_+$ domain, thus, the values $m = 2$ and $m = 0$ correspond to $N_+$ and $N_-$
vortices, respectively.

The true quasiparticle wavefunction $\psi(k, \theta_k)$ must be, of course, a single-value
function of $\theta_k$. This requirement imposes a certain quantization rule:
the value of $\mu + m / 2$ must be an integer number. Since $m$ is even for the
singly-quantized vortex, the angular momentum $\mu$ is an integer (cf. Ref.~\cite{refVolovikMajorana}).
 It means that defect does not change the
quantization rule and does not shift the zero energy level.

\subsection{Boundary conditions}

Now we need to rewrite the boundary condition~(\ref{eqBoundaryExact}) imposed on the quasiparticle wavefunction
in the quasiclassical form. Using the expressions~(\ref{eqQuasiClassicSubstitute})
and~(\ref{eqMomentumSubstitute}) we obtain:
\begin{equation}
	\label{eqBoundary1}
	\int\limits_0^{2\pi} e^{i k_F R \cos \alpha + i \mu \alpha}
	e^{i m \tau_3 \alpha/ 2}\widetilde \psi
	\left( R \cos \alpha \right)\; d\theta_k = 0 \ .
\end{equation}
Supposing the argument of the first exponent to vary rather fast  we can use
the stationary phase method in order to evaluate the integral.
The stationary phase points are given by the equation $\sin \alpha = \mu / k_F R = -b/ R$. This
equation has no solutions if the impact parameter is greater than the defect radius. It means that
the integral~(\ref{eqBoundary1}) is negligible and no boundary
contidions are required because the trajectory does not hit the defect.
In the opposite case $|b|<R$ there are two stationary
angles $\alpha_1 = -\arcsin (b/R)$ and $\alpha_2 = \pi - \alpha_1 $ that correspond to
the incident and reflected classical trajectories. The sum of the two
contributions provides the boundary condition:
\begin{equation}
	\label{eqBoundary2}
	e^{i\varphi_0}\widetilde \psi(s_0) = e^{-i \varphi_0}\widetilde \psi(-s_0) \ ,
\end{equation}
where $s_0 = \sqrt{R^2 - b^2}$ and
\begin{equation*}
	\varphi_0 = k_F s_0 + (\mu + m \tau_3 / 2) (\alpha_1 - \alpha_2)/2 + \pi / 4 \ .
\end{equation*}

\section{Excitation spectrum}
\label{secExcitationSpectrum}

\subsection{Large impact parameters $|b| > R$}

In this case the trajectory does not hit the defect and does not experience any reflection, thus,
the spectrum should be described by the standard CdGM
solution \cite{refCdGM}.

Here we find spectrum and wave functions  considering the imaginary part of $\widetilde D$ as
a perturbation \cite{VolovikCdGM}.
Neglecting the corresponding part in~(\ref{eqBdGPType1}) we find:
\begin{equation}
	-i \xi \tau_3 \frac{\partial \widetilde \psi}{\partial s} + \tau_1 \Re \widetilde D_b(s)
	\widetilde \psi = \varepsilon \widetilde\psi \ .
\end{equation}
This equation has a zero eigenvalue with the following eigenfunction:
\begin{equation}
	\widetilde \psi_{b} = \frac{1}{\sqrt{2I_b}} \left(
	\begin{split}
		i\\
		1
	\end{split}
	\right) e^{-K_b(s)} \ ,
\end{equation}
where
\begin{equation*}
	K_b(s) = \frac{1}{\xi} \int\limits_0^s \Re \widetilde D_b(s')\;ds',~
	I_b = \int\limits_{-\infty}^{+\infty} e^{-2K_b(s)}\;ds \ .
\end{equation*}
The first order perturbation theory yields the following excitation spectrum:
\begin{equation}
	\label{eqExcitationLargeMomentum}
	\varepsilon_b = \frac{1}{I_b}\int\limits_{-\infty}^{+\infty} \Im \widetilde D_b(s) e^{-2K_b(s)}\;ds \ .
\end{equation}

\subsection{Small impact parameters $|b| < R$}
In this case the quasiparticle experiences reflection from the cylinder
surface which modifies the spectrum. In order to solve the
equation~(\ref{eqBdGPType1}) we have to take into account the boundary
conditions~(\ref{eqBoundary2}), so we introduce the function:
\begin{equation}
	\Psi(s) = \left\{
		\begin{split}
			e^{+i\varphi_0}\widetilde \psi(s), s > 0\\
			e^{-i\varphi_0}\widetilde \psi(s), s < 0\\
		\end{split}
	\right. \ .
\end{equation}
Due to the boundary condition~(\ref{eqBoundary2}) $\Psi(s_0) = \Psi(-s_0)$. The new function
satisfies the following equation:
\begin{equation}
	\label{eqBdGQuasiClassic1}
	-i\xi \tau_3 \frac{\partial \Psi}{\partial s} + \left(\tau_+ G_b(s) + h.c.\right) \Psi
	= \varepsilon \Psi \ ,
\end{equation}
where $G_b(s) = e^{i \phi \sign s} \widetilde D_b(s)$, $\phi = m(\alpha_1 - \alpha_2) / 2$.
The equation~(\ref{eqBdGQuasiClassic1}) is similar to a quasiclassical equation descrribing a
Josephson junction: the order parameter is constant if $s \to \pm \infty$. Assuming such step like
form of
the order parameter along the trajectory we find the energy \cite{refBeenakkerVanHoutenPRL}:
\begin{equation}
	\label{eqEnergyJosephson}
	\varepsilon = \chi \cos (\phi + \pi/2) = -\chi \sin \phi \ ,
\end{equation}
where $\chi = \sign\left(\cos \phi\right)$.
The energy depends only on the order parameter phase difference on the
trajectory ends.
The additional phase difference $\pi$ arises from the order parameter symmetry property $G_b(s) =
-G_b^\ast(s)$.
The above approximate solution can be, of course, improved if we take account of the Doppler shift of quasiparticle
energy caused by the superflow around the core. Such improvement is particularly important for the case of
$N_-$ vortex when the expression~(\ref{eqEnergyJosephson}) yields
$\varepsilon = 0$ for all impact parameters and, thus, does not allow to get the correct slope of the anomalous spectral branch.

\begin{figure*}[t]
	\center
	\subfigure[$N_+$ vortex]{
		\includegraphics[width=0.48\linewidth]{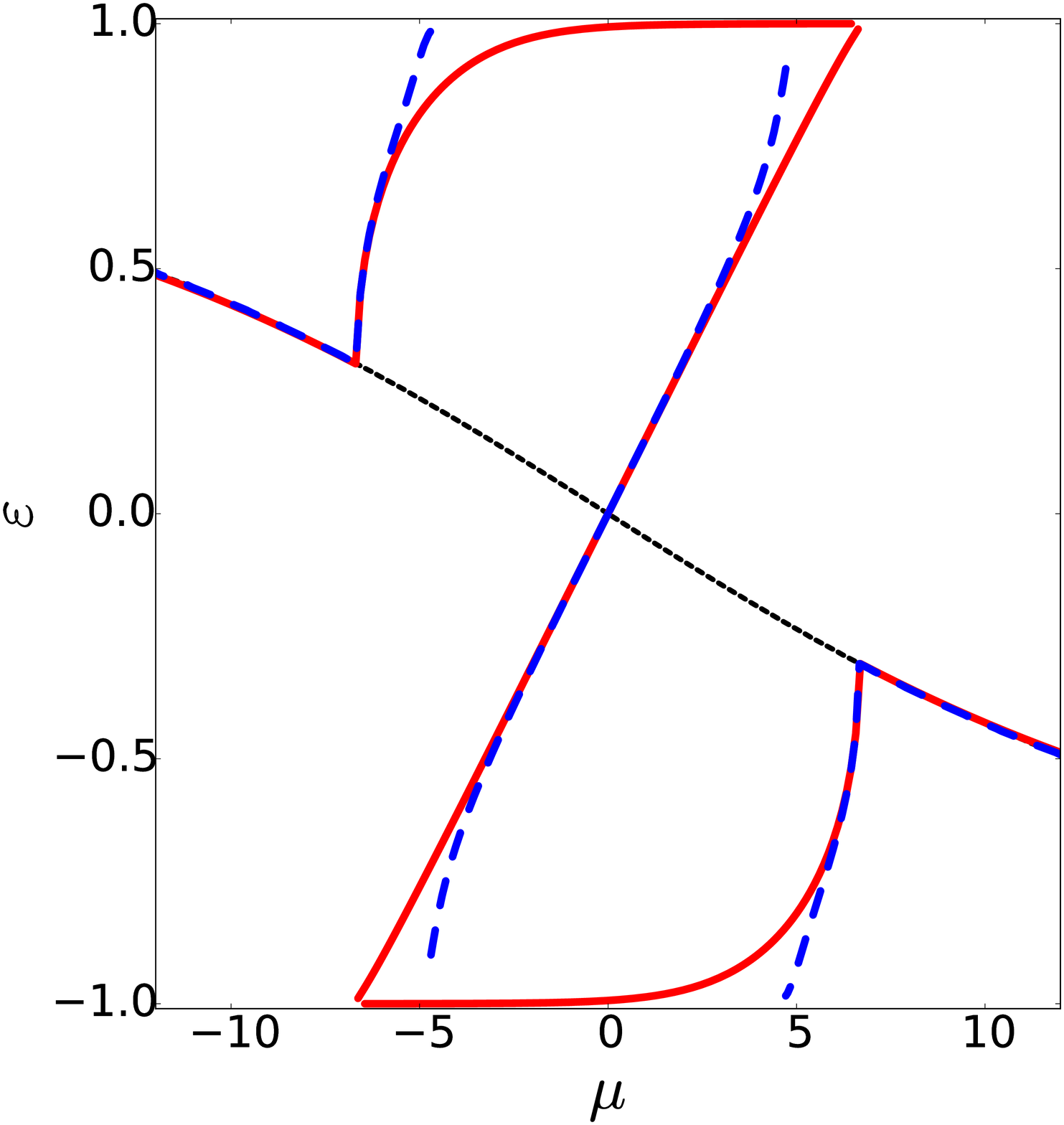}
	}
	\subfigure[$N_-$ vortex]{
		\includegraphics[width=0.48\linewidth]{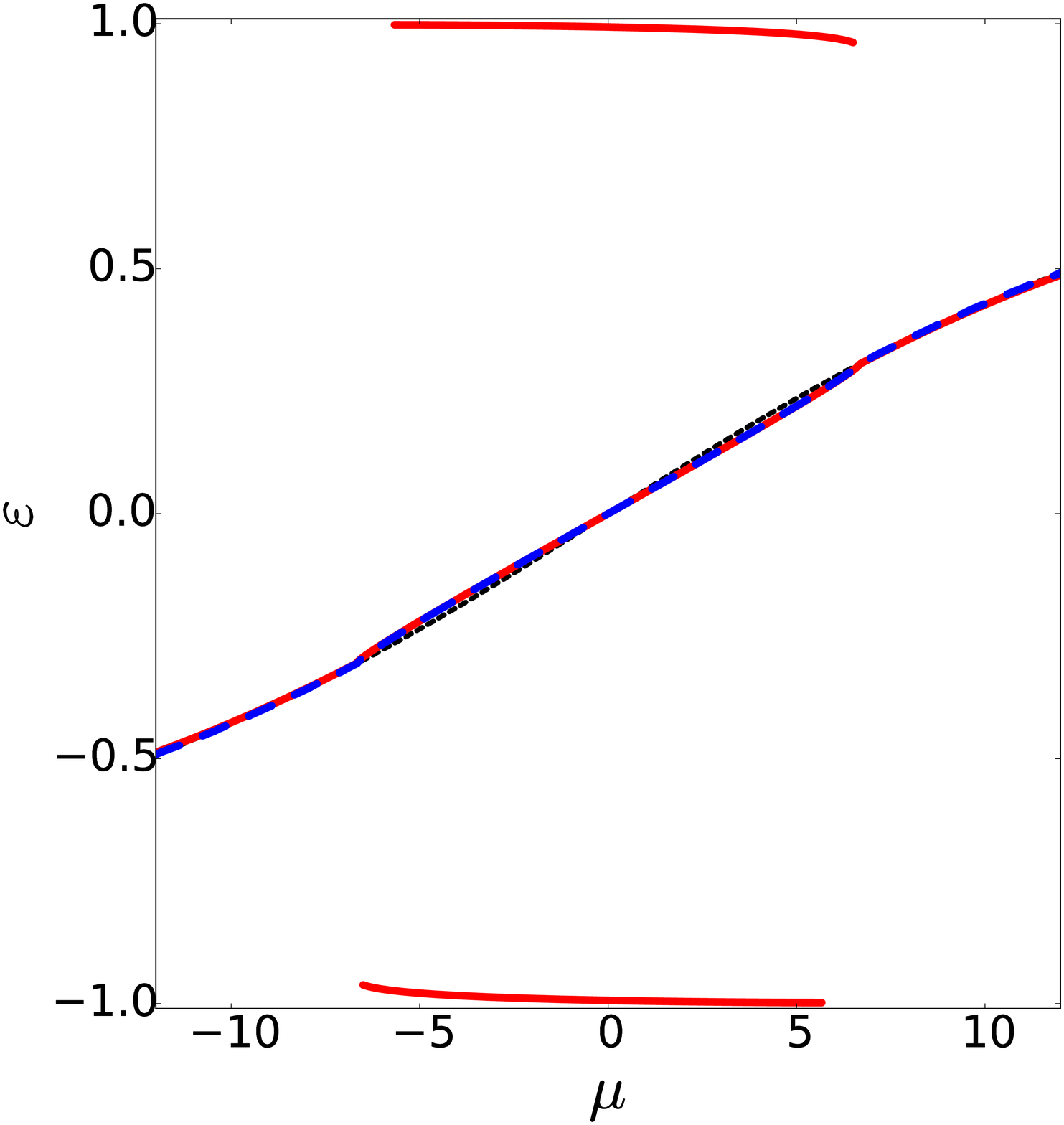}
	}
	\caption{Quasiparticle spectrum for two vortex types found from
the solution of the quasiclassical
		equation~(\ref{eqBdGPType1}). The numerical solution is shown by the
		solid red lines, the dashed blue lines correspond to the results of
		perturbation theory, the black
		dashed line is the CdGM branch. The defect radius is $R = 0.4\xi$.}
	\label{figPTypeQuasiclassicSpectrum}
\end{figure*}

We can apply the perturbation theory used above in order to obtain a more precise solution.
First we neglect
the imaginary part of $G$ and obtain the wave functions corresponding to the zero energy $\varepsilon = 0$:
\begin{equation}
	\label{eqWaveFunctionZeroOrder}
	\Psi_b(s) = \frac{1}{\sqrt{2 I_b}} \left(
	\begin{split}
		i \chi\\
		1
	\end{split}
	\right) e^{-K_b(s)},~s > s_0 \ ,
\end{equation}
where
\begin{equation*}
	K_b(s) = \frac{\chi}{\xi}\int\limits_{s_0}^s \Re G_b(s')\;ds',~
	I_b(s) = 2 \int\limits_{s_0}^{+\infty} e^{-2K_b(s)}\;ds \ .
\end{equation*}
The eigenfunction is even, $\Psi_b(s) = \Psi_b(-s)$. This localized solution can be used as a
zero-order approximation for the wave function. Within the first-order perturbation theory we
find the spectrum:
\begin{equation}
	\label{eqExcitationSmallMomentum}
	\varepsilon_b = \frac{2 \chi}{I_b} \int\limits_{s_0}^{+\infty} \Im G_b(s) e^{-2K_b(s)}\; ds \ .
\end{equation}
The behavior of the subgap spectral branches found within this perturbation procedure is illustrated
in Fig.~\ref{figPTypeQuasiclassicSpectrum}(a) and (b) for  $N_+$ and $N_-$ vortices, respectively.
To verify the approximate solution we have also solved the quasiclassical
equations~(\ref{eqBdGPType1}) numerically.
The results of numerical calculations shown in Fig.~\ref{figPTypeQuasiclassicSpectrum} demonstrate a good coincidence
 with the ones obtained using the perturbation approach except the energies close to the superconducting gap $\Delta_0$.
 The failure of the pertubation procedure in this limit arises
 from the divergence of the wave function~(\ref{eqWaveFunctionZeroOrder}) localization radius.

The spectrum of the $N_-$ vortex only slightly differs from
the CdGM solution (see Fig.~\ref{figPTypeQuasiclassicSpectrum}(b)). In this vortex the order parameter vorticity in $\mathbf r$-space
is compensated by its chirality in $\mathbf k$-space and the phase difference at the
ends of every classical trajectory is always equal to $\pi$. The defect effectively
changes the order parameter amplitude along the trajectory and modifies the spectrum.

In opposite, for the $N_+$ vortex the phase difference at the
ends of classical trajectory causes a significant spectrum modification even for small impact parameters (see Fig.~\ref{figPTypeQuasiclassicSpectrum}(a)).
As a result, the subgap spectrum consists of three branches. Within the perturbation approach
these branches reveal themselves in the spectrum discontinuity
 at the points $b = \pm R / \sqrt{2}$, where perturbation theory
is not applicable.
One can observe this energy discontinuity even in the simplified
expression~(\ref{eqEnergyJosephson}) where
$\chi$ changes sign at the points $b = \pm R / \sqrt{2}$.
\begin{figure}[th]
	\center
	\includegraphics[width=\linewidth]{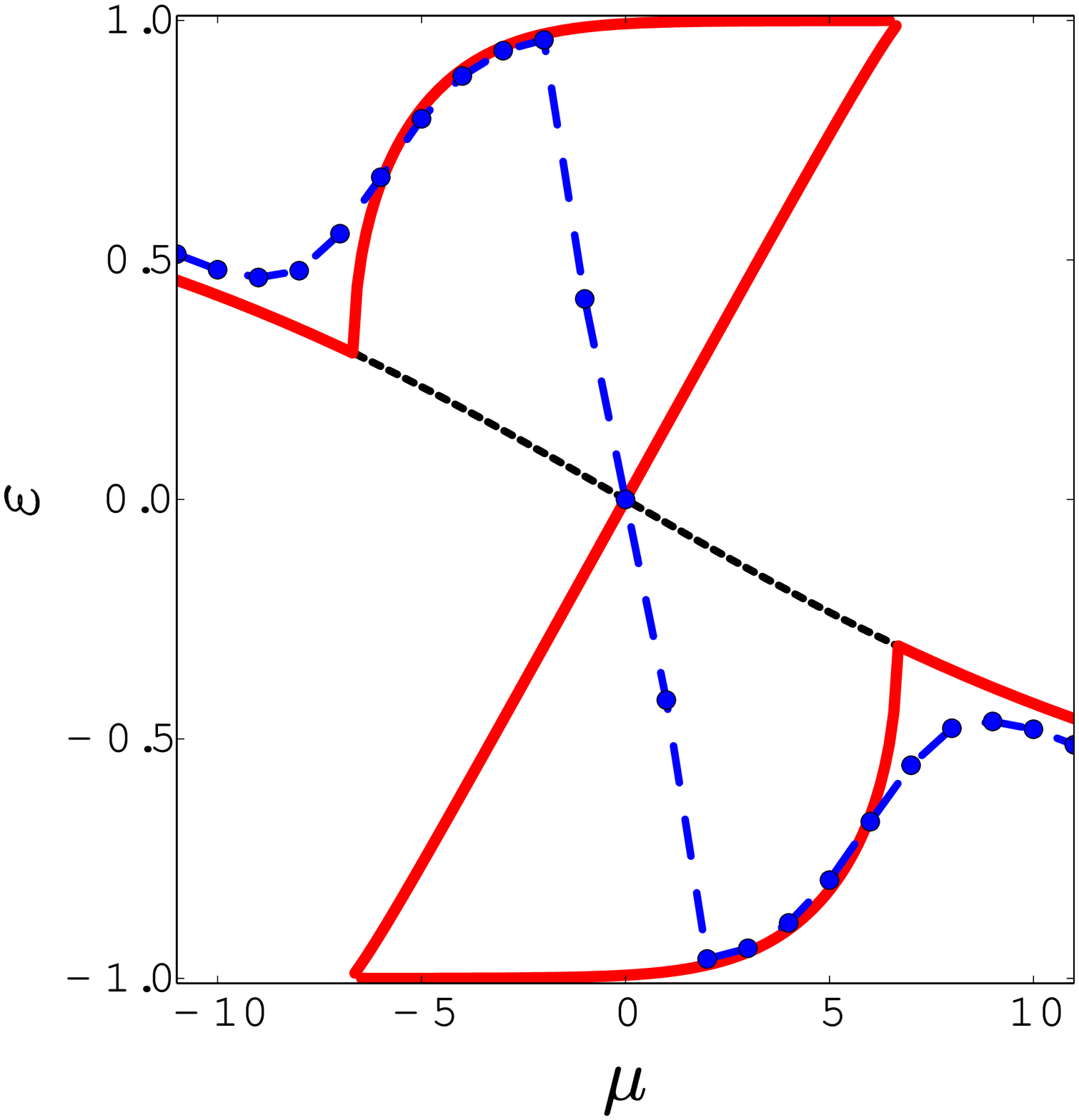}
	\caption{Comparison of the subgap spectral branches (red lines) in the $N_+$ vortex
with the ones found in
Ref.~\cite{refRosensteinShapiroJOP} from numerical simulations on the basis of BdG theory
(blue dashed lines).}
	\label{figPTypeNumericExactSpectrum}
\end{figure}
There are two branches which transform into the CdGM branch at large $|b| > R$
and approach the superconducting gap at small $b$.
The similar spectral branches have been observed earlier
in the spectrum of a pinned vortex in a $s$-wave superconductor \cite{MelnikovSamokhvalovPhysRevB79p134529}.
In addition to these branches there is an almost linear branch that goes
through the origin with the slope inversed with respect to the CdGM solution (cf. the introductory section).
We propose that this branch
corresponds to the edge states bound to the surface of the unconventional
superconductor. The spectrum of these surface states can be
easily found within the quasiclassical approach solving the equation~(\ref{eqBdGQuasiClassic1}) with
$D = \exp(i \theta_k)$ that corresponds to the homogeneous chiral domain. Performing the same calculations
as we had done for a vortex, we will obtain the following spectrum:
\begin{equation}
	\label{eqHoleSpectrum}
	\varepsilon_b = \left\{
		\begin{split}
			-\frac{b}{R},~&|b| <  R\\
			-\sign b,~&|b| >  R
		\end{split}
	\right. \ .
\end{equation}
This quasiparticle spectrum is very close to the anomalous spectrum branch in the
Fig.~\ref{figPTypeQuasiclassicSpectrum}(a), so we can claim that this branch corresponds
to the surface states.

This anomalous branch with the inversed slope has been overlooked by the authors of Ref.~\cite{refRosensteinShapiroJOP}. The
results of their numerical calculations are shown in the
Fig.~\ref{figPTypeNumericExactSpectrum}. Note that at large angular momenta the spectrum found in Ref.~\cite{refRosensteinShapiroJOP}
transforms into the CdGM solution as expected and is close
to our numerical solution of~(\ref{eqBdGPType1}).

\section{Local density of states}
\label{secLDOS}

As a next step we turn to the calculations of the local density of states (LDOS) which can be probed, e.g.,
in the  STM/STS studies.
The measurable quantity in these experiments is the local differential conductance:
\begin{equation}
	\label{eqLocalConductance}
	\frac{dI/dV}{(dI/dV)_N} = \int\limits_{-\infty}^{+\infty} \frac{N(\mathbf r, \epsilon)}{N_0}
	\frac{\partial f(\epsilon - eV)}{\partial V}\;d\epsilon \ ,
\end{equation}
where $V$ is the voltage, $(dI/dV)_N$ is the junction conductance in the normal state, $N$ is the LDOS in the superconductor,
$N_0$ is the normal state DOS and $f(\epsilon) =
\left[1 + \exp(\epsilon/T)\right]^{-1}$ is the Fermi function.
Within the quasiclassical approach
the local DOS is determined as follows:
\begin{equation}
	\label{eqLDOS}
	N(\mathbf r, \epsilon) = \frac{1}{2\pi} \int k_F \left|u_{b}(\mathbf r)\right|^2
	\delta\left(\epsilon - \epsilon_{b}\right)\;db \ .
\end{equation}
Substituting~(\ref{eqLDOS}) into~(\ref{eqLocalConductance}) we
obtain:
\begin{equation}
	\frac{dI/dV}{(dI/dV)_N} = k_F \int\limits_{-\infty}^{+\infty} \frac{|u_b(\mathbf r)|^2}{N_0}
	\frac{\partial f(\epsilon_b - eV)}{\partial V}\;db \ .
\end{equation}


The local DOS and the differential conductance are both expressed
through the electron-like wave function $u_{b}(\mathbf r)$ corresponding to
the energy $\epsilon_{b}$.
We use the expressions~(\ref{eqQuasiClassicSubstitute}), (\ref{eqAngularEikonal}) and~(\ref{eqMomentumSubstitute})
in order to restore $u_b(r)$. If
$k_F r \gg 1$ it can be evaluated using the stationary phase method.
In this limit the wave function is determined by the quasiclassical wave functions at
two classical trajectories passing through the point $(r,\theta)$:
\begin{equation}
	u_{b}(r, \theta) = \frac{e^{i \mu \theta + i m \theta / 2}}{\sqrt{2\pi i k_F r}}
	\sum\limits_{j=1,2} e^{i \varphi_j}
	\frac{\widetilde \psi_{u,b}\left(r \cos \beta_j\right)}{\sqrt{\cos \beta_j}} \ ,
\end{equation}
where $\varphi_j = k_F \cos \beta_j + i (\mu + m/2) \beta_j$, $\beta_1 = -\arcsin b/R$, $\beta_2 = \pi
- \beta_1$. Neglecting the part oscillating at the atomic length scale and applying the normalization condition we obtain:
\begin{equation}
	\left|u_{b}(r)\right|^2 = \frac{\exp\left[-2K_b\left(\sqrt{r^2 - b^2}\right)\right]}
	{\pi I_b \sqrt{r^2 - b^2}}
\end{equation}

\begin{figure*}[t]
	\centering
	\subfigure[$N_+$ vortex] {
		\includegraphics[width=0.48\linewidth]{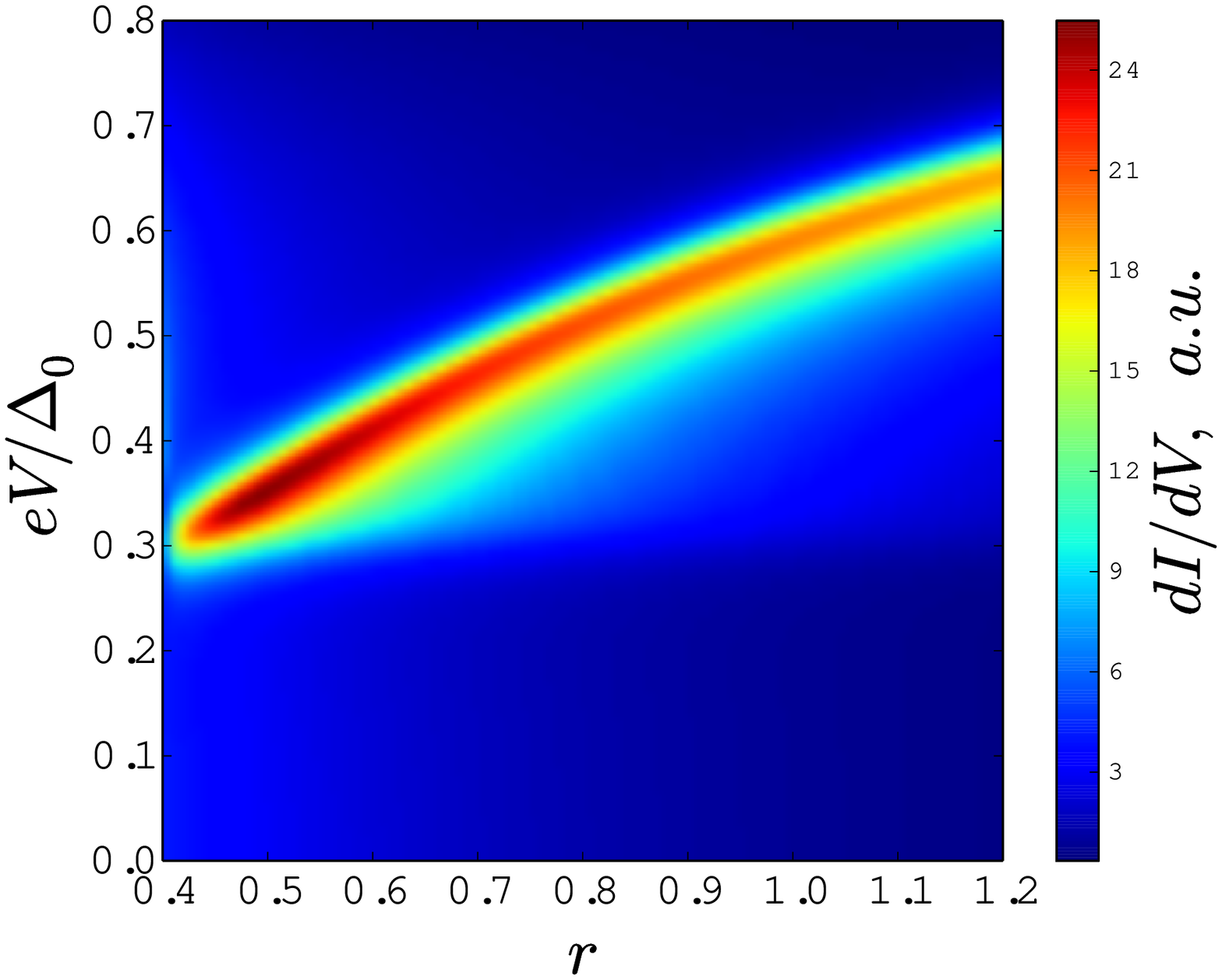}
	}
	\subfigure[$N_-$ vortex] {
		\includegraphics[width=0.48\linewidth]{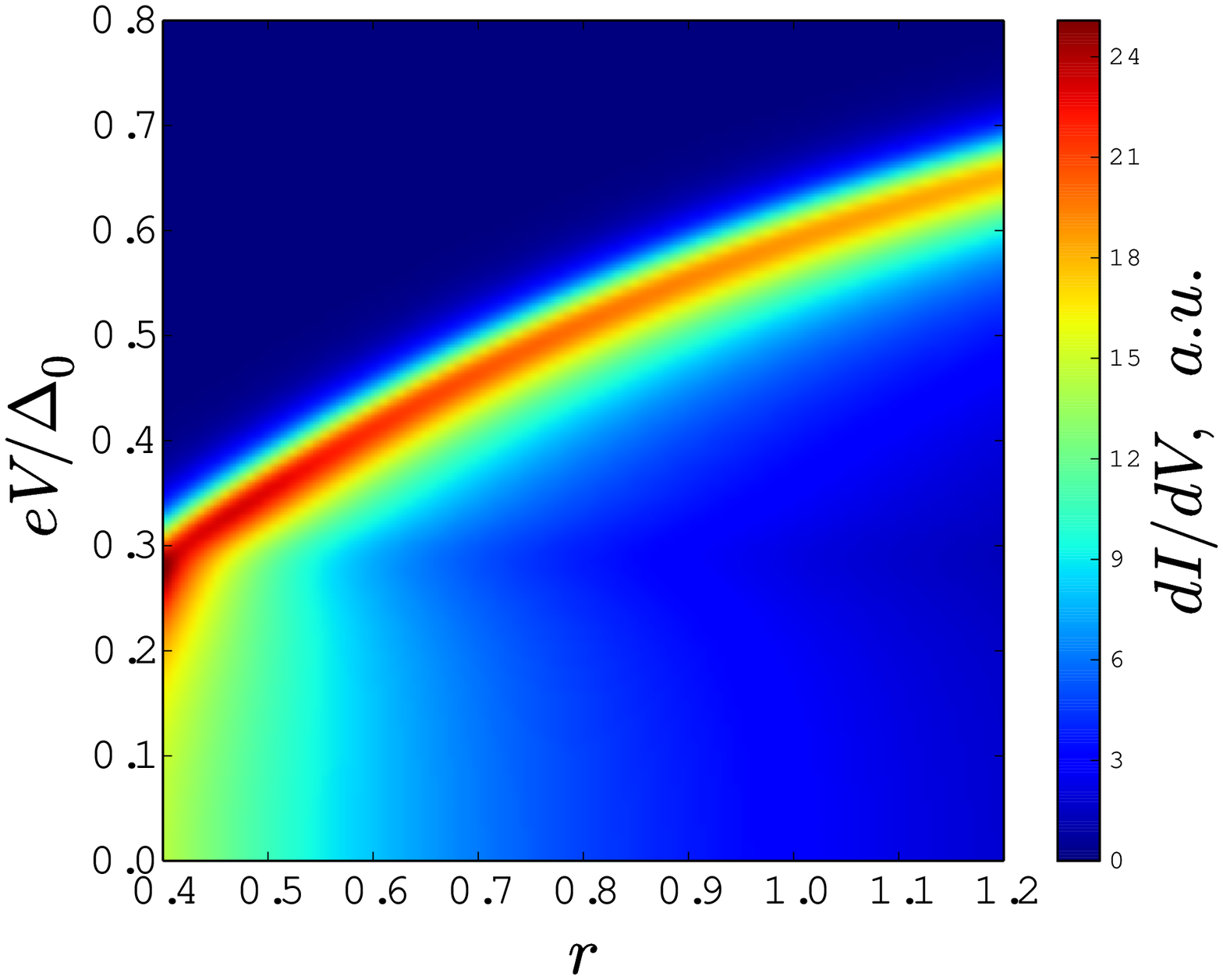}
	}
	\caption{The local differential conductance vs the voltage $V$ and the distance
	from the vortex axis $r$ for different vortex types. Here we put $R = 0.4\xi$, $T
	= 0.02 \Delta_0$.}\label{figLDOS}
\end{figure*}

The local conductance is shown in the Fig.~\ref{figLDOS} for different types of vortices.
The conductance profile for the $N_-$ vortex (Fig.~\ref{figLDOS} b) reveals
the typical CdGM behavior for $r>R$ \cite{refHessRobinsonZeroBiasPRL}.
This conclusion is no more valid if we consider $N_+$ vortex where the large slope of the inversed anomalous branch  causes
strong changes in the LDOS pattern (Fig.~\ref{figLDOS} a). The local conductance distribution in this case is similar to the
one for a pinned vortex
in the $s$-wave superconductor \cite{MelnikovSamokhvalovPhysRevB79p134529}.


\section{High-frequency conductivity}
\label{secHallConductivity}

Besides the STM/STS studies there exists another efficient method for experimental investigation
of quasiparticle subgap spectrum based on the measurements of the conductivity
tensor at finite frequencies.
In the classical limit the interaction of the quasiparticles with the
high-frequency field can be described using the following Hamiltonian:
\begin{equation}
	\label{eqHamiltonian}
	H(\mu, \theta) = \epsilon(\mu) + \hbar \mathbf {k_F} \mathbf {v_s} \ ,
\end{equation}
where $\epsilon(\mu)$ is the energy of the anomalous spectral branch and
$\mathbf {v_s}$ is the superfluid velocity induced by the electromagnetic
field. In London limit it is proportional to the vector potential
$\mathbf{v_s} = Q \mathbf A$. Taking a circularly polarized field $\mathbf {v_s}$ with
frequency $\Omega$, finally we obtain the following Hamiltonian:
\begin{equation}
	\label{eqHamiltonian2}
	H(\mu, \theta) = \epsilon(\mu) + 2\hbar Q k_F Re \left(A_\pm e^{\pm i \theta - i
	\Omega t}\right) \ ,
\end{equation}
where the sign ``+'' or ``-'' denotes the circular polarization orientation and
$A_\pm$ is the complex magnitude, i.e. the total magnetic potential is
$\mathbf A =Re\left(e^{-i\Omega t} A_\pm (\mathbf x_0 \pm i \mathbf y_0)\right)$. In order to find conductivity one should
solve the Boltzmann equation written for the quasiparticle distribution function:
\begin{equation}
	\frac{\partial f}{\partial t} + \frac{1}{\hbar}\left(\frac{\partial H}{\partial \mu}
	\frac{\partial f}{\partial \theta} - \frac{\partial H}{\partial
	\theta} \frac{\partial f}{\partial \mu}\right) = - \nu\left(f - f_0\right) ,
\end{equation}
where $f(\theta, \mu, t)$ is the distribution function, $f_0 (\mu)$ is the
equilibrium distribution function and $\nu$ is the quasiparticle relaxation rate.
This equation can be solved within the perturbational approach, so that the
total distribution function is represented as sum $f = f_0 + f_1$, where $f_1$
is the first-order perturbation term:
\begin{equation}
	f_1(\theta, \mu, t) = 2 Re \frac{\pm i c Q k_F E_\pm}{\Omega \left (\Omega \mp \omega -
	i \nu\right )} \frac{\partial f_0}{\partial \mu} e^{\pm i
	\theta - i \Omega t} ,
\end{equation}
where $\hbar \omega = \partial \epsilon / \partial \mu$ and $E_\pm = i
\Omega/c A_\pm$ is the electric field complex magnitude. Let us find the
current for the zero-temperature case:
\begin{equation}
	\label{eqCurrent}
	j_\pm = \frac{e c Q k_F}{2 \Omega \left (i \Omega \mp i \omega_0
	+ \nu \right)} E_\pm \ ,
\end{equation}
where $\hbar \omega_0 = \partial \epsilon / \partial \mu |_{\mu = 0}$. One can easily obtain Ohmic and Hall conductivities from
the~Eq.(\ref{eqCurrent}):
\begin{gather}
	\sigma_O = \frac{ecQ}{\Omega} \frac{\nu + i\Omega}{\left(\nu + i
	\Omega\right)^2 + \omega_0^2}\\
	\sigma_H = -\frac{ecQ}{\Omega} \frac{\omega_0}{\left(\nu + i
	\Omega\right)^2 + \omega_0^2}
	\label{eqHallConductivity}
\end{gather}
So one can see that the sign and the value of the Hall conductivity are
strongly determined by the
slope of the anomalous spectral branch at the Fermi level. The Hall
conductivity can be probed experimentally, one of the methods is the polar
Kerr effect measurements \cite{refXiaMaenoPhysRevLett}. The following
experiment can be proposed: the Hall conductivity can be measured at zero
field and for the two opposite orientations of the magnetic field. According
to the~Eq.(\ref{eqHallConductivity}) the Hall conductivity for all three
cases has the same sign. Since the slope of the spectral branch for the $N_+$
vortex almost coincides with the one without vortex the Hall
conductivity for this field orientation should coincide with the zero field Hall conductivity. 
Note that such coincidence is a fingerprint of the quasiparticle spectrum calculated above comparing to the one found in Ref.~\cite{refRosensteinShapiroJOP}.
For the opposite field orientation the Hall
conductivity appears to be suppressed due to the small spectral branch slope.
Certainly, the above picture for the $N_+$ vortex is valid for rather low frequencies $\Omega<\Delta_0 R/\xi$, i.e., when the rf field can not induce
transitions to the levels at the broken CdGM branch in Fig.\ref{figPTypeNumericExactSpectrum}. 

\section{Summary}
\label{secSummary}

We have calculated the excitation spectrum in vortices pinned by columnar defects
in chiral $p$ wave superconductors.
The spectrum is shown to depend strongly on the orientation of the magnetic field
with respect to the internal angular momentum (chirality) of the Cooper pairs.
If the magnetic field produce flux lines with the vorticity opposite to this internal angular momentum
the quasiparticle spectra in pinned vortices are only slightly disturbed by the presence of defects.
In the case of coinciding signs of vorticity and chirality the subgap spectra in pinned vortex cores appear to be
strongly different from the ones for free vortices: the anomalous branch at small impact parameters changes its slope
resulting in the change in the LDOS pattern and contribution of the
quasiparticles into the Ohmic and Hall conductivities at finite frequencies.

%

We thank A. Samokhvalov
for stimulating discussions and G. Volovik for valuable comments.
This work was supported by the Dynasty Foundation, Russian Foundation for
Basic Research and the grant of the Russian Ministry of Science and Education $(02.\text{B}.49.21.0003)$.
\bibliographystyle{apsrev}
\bibliography{bibliography}

\end{document}